\documentclass{aastex}
\usepackage{amsbsy,emulateapj5,epsfig,onecolfloat,apjfonts}   

\shorttitle{Tangential velocities from Gravitational Lensing}
\shortauthors{S.~M.~Molnar, M.~Birkinshaw}

\newcommand{\bfalpha}{\boldsymbol{\alpha}}
\newcommand{\bfbeta}{\boldsymbol{\beta}}
\newcommand{\bftheta}{\boldsymbol{\theta}}
\newcommand{\bfdelta}{\boldsymbol{\delta}}
\newcommand{\bfgamma}{\boldsymbol{\gamma}}

\begin{document}

\twocolumn
[
\title{Determining Tangential Peculiar Velocities of Clusters of Galaxies using \\
                  Gravitational Lensing}

\author{
Sandor M. Molnar\altaffilmark{1}, and Mark Birkinshaw\altaffilmark{2}
}

\begin{abstract}
We propose two new methods for measuring tangential peculiar velocities of 
rich clusters of galaxies. Our first method is based on weak gravitational 
lensing and takes advantage of the differing images of 
background galaxies caused by moving and stationary gravitational potentials. 
Our second method is based on measuring relative frequency 
shifts between multiple images of a single strongly lensed background galaxy. 
We illustrate this method using the example of galaxy cluster CL~0024+1654.
\end{abstract}

]

\altaffiltext{1}{Department of Physics and Astronomy, Rutgers University, 
136 Frelinghuysen Road, Piscataway, NJ~08854}

\altaffiltext{2}{Department of Physics, Bristol University, Tyndall Avenue, 
Bristol, BS8 1TL, UK}

\section{Introduction}
\label{s:Introduction}

The distribution of peculiar velocities of galaxies and clusters of galaxies 
is sensitive to the overall 
matter density in the Universe on large scales, and can be used to study 
structure formation. Peculiar velocities of galaxies have been used to
reconstruct matter density fields 
(Colberg et al. 2000; Dekel \& Ostriker 1999; Peacock 1999). 
Peculiar velocities of clusters could be used to determine large-scale bulk 
motion \citep{KashAtri00} and check the reconstructed matter density fields. 
Direct methods to measure peculiar velocities of clusters of galaxies
suggested so far are based on the frequency changes that moving clusters
generate in the background photons of the CMBR due to inverse Compton
scattering or gravitational lensing.

Measurements of cluster radial peculiar velocities are based on
inverse Compton scattering of CMBR photons off hot electrons in the 
intra-cluster gas, the Sunyaev-Zel'dovich effect \citep{SunyZe80}. 
On average the CMBR photons gain energy from this scattering and
a static and stationary cluster induces a 
decrement in the Rayleigh-Jeans part of the spectrum (the static
thermal SZ effect). If the cluster is moving relative to the
CMBR, an additional kinematic SZ effect is generated because
of the bulk motion of the intra-cluster gas.
The ratio of the kinematic to static thermal SZ effect,
$\Delta T_{KSZ}/\Delta T_{SSZ}$, in the Rayleigh-Jeans spectral region 
may be expressed as

\begin{equation} \label{E:restframe}
  { \Delta T_{KSZ} \over \Delta T_{SSZ} } = 0.085 
                            \;  \Biggl[ {v_r \over 1000\;{\rm km\;s^{-1}}} \Biggr]
                            \;  \Biggl[ {k T_e \over 10\;{\rm keV}} \Biggr]^{-1}
, 
\end{equation}
where $v_r$ is the radial peculiar velocity, $T_e$ is the electron temperature, 
and $k$ is the Boltzmann constant, so that in moderate- or high-temperature 
clusters of galaxies the static effect will dominate. 
The kinematic SZ effect can be separated from 
the static effect using its different frequency distribution, even though 
their spatial distribution is the same 
(for recent reviews see Birkinshaw 1999 and Rephaeli 1995).
The accuracy to which this radial velocity effect can be measured in a 
cosmological context was analyzed by Aghanim, Górski \& Puget (2001). 
At present only upper limits exist on radial velocities of clusters based on 
the kinematic SZ effect. 
Holzapfel et al. (1997) determined limits of peculiar radial velocities
for two clusters, A2163 and A1689, with $\pm 1 \sigma$ ranges: 
$-390\; \rm km\;s^{-1} \le v_r \le$
\\
$1860\; \rm km\;s^{-1}$ and 
$-460\; \rm km\;s^{-1} \le v_r \le 985\; \rm km\;s^{-1}$.

Tangential peculiar velocities of clusters can be determined using the
moving cluster effect (Birkinshaw \& Gull 1983; see also 
Pyne \& Birkinshaw 1993 and Gurvits \& Mitrofanov 1986).
This effect is due to the fact that gravitational bending of 
light deforms the dipolar CMBR temperature anisotropy in the rest
frame of the cluster, and thus when this pattern is transformed 
back to the CMBR rest frame (the observer's frame) 
we do not recover the original, uniform, CMBR distribution. 
This effect is only of order $5 \ \rm \mu K$, far less than the static or
kinematic SZ effects from prominent clusters (400 $\mu \rm K$ and 50 $\mu \rm K$). 
Measurements of such small fluctuations in the CMBR on
the few arcminute scale of clusters is beyond our current
capability, and so no attempt has yet been made to measure this
effect. Sensitivity improvements of a factor $\sim 100$ are
needed if this effect is to be measured, and this will be
difficult in the presence of primary CMBR structures with
the same spectrum.

In this letter, we propose two new methods for measuring tangential peculiar 
velocities of rich clusters of galaxies without recourse to CMBR observations.
Our first method is based on weak gravitational lensing.
The distortions of images of background objects caused by a moving gravitational 
lens are different from those caused by a static lens.
We calculate the resulting weak lensing
signal and discuss how this effect could be used to measure tangential
velocities of clusters using current technology. 
Our second method is based on frequency shifts generated by moving gravitational 
lenses, but instead of using the CMBR as a reference, we suggest 
using multiple images of a single strongly lensed background galaxy. 
We illustrate this second method using the example of galaxy cluster CL 0024+1654.

\section{Theory}
\label{s:Theory}

In the following derivation we use the thin lens approximation for gravitational 
lensing and assume that the deflection angle is small ($\ll 1$ radian).
This approximation is well justified for gravitational lensing caused by most
clusters of galaxies (see for example Schneider, Ehlers \& Falco 1992).
In this approximation, the lens equation becomes

\begin{equation} \label{E:thinlens}
  {\bftheta} = {\bftheta}^S + {\bfalpha}
, 
\end{equation}
where ${\bftheta}^S$ and ${\bftheta}$ are the source and image coordinates, and
${\bfalpha}$ is the observer's frame bend angle \citep{Schnet92}. 
We choose coordinate axes $\hat{\bf x}$ and $\hat{\bf y}$ to lie in the plane of 
the sky. In the rest frame of the cluster, the bending angle is 

\begin{equation} \label{E:restframe}
  {\bfdelta}(x,y) \equiv (\delta_x,\delta_y) = \nabla \psi(x,y)
, 
\end{equation}
where $x$ and $y$ are angular coordinates in the sky 
and $\psi(x,y)$ is the 2-dimensional gravitational deflection potential.
The change in the photon four-momentum in the observer's frame is 

\begin{equation} \label{E:Delta}
  \Delta = \Bigl[ 
     {\cal L}^{-1}(\bfbeta) {\cal R}(\bfdelta) {\cal L}(\bfbeta) - {\cal I} \Bigr]
                          \gamma^{in} = {\cal D} \, \gamma^{in}
,
\end{equation}
where ${\cal L}$, ${\cal R}$, ${\cal I}$ and ${\cal D}$, are the four-dimensional 
Lorentz transformation, rotation (caused by gravitational lensing), identity 
and distortion matrices, $\gamma^{in}$ is the incoming photon four-momentum,
and ${\bfbeta} \equiv {\bf v}/c$ is the peculiar velocity of the moving lens
in units of the speed of light, $c$.

First we briefly discuss the case when the gravitational lens is moving radially, 
i.e. ${\beta} \equiv v_R /c$.
In this case, for small bending angles and peculiar velocities 
($\delta \ll 1$ and $ \beta \ll 1$), from Equation~(\ref{E:Delta}), 
we obtain the deflection angle 

\begin{equation}  \label{E:alfa_rad}
 \alpha = \bigl(1 + {\beta} \bigr) \delta
.
\end{equation}
This result agrees with the results of Frittelli, Kling \& Newman (2002; Equation 99).

In theory this correction to the bending angle should be taken into account
when using weak lensing to determine masses of clusters of galaxies.
Cluster masses are determined from the bending angle.
If the lens is moving radially, the bending angle 
changes because of its radial velocity, and the correct
cluster mass, $M$, is related to the mass determined by
ignoring the radial motion of the lens, $M_0$, by

\begin{equation}
   M = {M_0 \over 1 + \beta} \quad .
\end{equation}
However, since we expect that the peculiar velocities of clusters are less than 
$3000\; \rm km\; s^{-1}$, this effect is less than one percent, and is negligible 
relative to other errors even if we take it into account that 
linear structure formation theories underestimate peculiar velocities of 
clusters by about 40$\%$ (see Colberg et al. 2000).
Therefore we do not discuss this case further.

We now turn to the effect caused by tangential peculiar
velocities. For simplicity, we assume that the peculiar velocity
of the gravitational lens, $\beta \equiv \nu_T/c$ is 
entirely in the plane of the sky.
We choose $\hat{\bf x}$ parallel to $\bfbeta$.
The bending angle in the observer's frame can be read from the spatial 
components of $\Delta$ (Equation~\ref{E:Delta},
 assuming $\delta \ll 1$ and $ \beta \ll 1$) as

\begin{equation} \label{E:alfa}
  \bfalpha = \left(
      \begin{array}{ll}
       \gamma  &  0 \\
       0       &  1
      \end{array}
\right) \bfdelta
,
\end{equation}
where $\gamma=1/\sqrt{1-\beta^2}$. 
This gives $\bfalpha = \bfdelta$ when $\beta=0$, as expected. 
The Jacobian, $\cal J$, of
the transformation from the source to the image plane can be
seen from Equations (\ref{E:thinlens}) and (\ref{E:alfa}) to be

\begin{equation}
         {\cal J} = {\cal I}^{2D} + {\cal D}^{2D}
,
\hspace*{5mm}        
\end{equation}
where ${\cal I}^{2D}$ is the two-dimensional identity matrix,
and the two-dimensional matrix

\begin{equation} \label{E:shear}
 {\cal D}^{2D} \equiv \left( \begin{array}{ll}
                             \gamma \psi_{xx} & \gamma \psi_{xy}            \\
                             \psi_{yx}        &        \psi_{yy} 
                              \end{array} \right)          
,              
\hspace*{5mm}        
\end{equation} 
(where $\psi_{ij}$ are second partial derivatives of the deflection potential with 
respect to variables $i$ and $j$) describes the distortion. 
The distortion matrix can be decomposed as 

\begin{equation} \label{E:decom}
 {\cal D}^{2D} = \left( \begin{array}{ll}
                      - \kappa - {\tilde \gamma_1} & {\tilde \gamma_2}             \\
                      {\tilde \gamma_2}            & - \kappa + {\tilde \gamma_1}
                        \end{array} \right)  + {\cal R}^{2D}(\varphi_0)
,
\hspace*{5mm}        
\end{equation} 
where the convergence, $\kappa$, and shear, $(\tilde{\gamma}_1,\tilde{\gamma}_2)$,
are determined by the deflection potential, $\psi$,
and the second term, ${\cal R}^{2D}(\varphi_0)$, is a rotation of the source image 
by an unmeasurable orientation angle $\varphi_0$, from the arbitrary assignment of 
axes $\hat{\bf x}$ and $\hat{\bf y}$  (see for example Peacock 1999). 
From Equations~(\ref{E:shear}) and (\ref{E:decom}) we conclude that the shear is characterized by
the two-dimensional vector field, $\tilde{\bfgamma}$, with components

\begin{eqnarray}  \label{E:gamma1}
  {\tilde \gamma_1} & = & {1 \over 2} \big[ -\gamma \psi_{xx} + \psi_{yy} \big]  \\
  {\tilde \gamma_2} & = & {1 \over 2} \big[ \gamma \psi_{xy} + \psi_{yx} \big]   
.
\end{eqnarray}
Whenever these second derivatives of the deflection potential exist, 
we should be able to determine the tangential velocity (which enters here through 
$\gamma$) and the deflection potential separately since the potential is a scalar, 
continuous field. 
Unless the deflection potential is such that 
$\partial^2 (\nabla^2 \psi)/ \partial x \partial y$ vanishes identically, 
$\gamma$ can be determined as

\begin{equation}  \label{E:gamma2}
  \gamma = {  {\tilde \gamma_{2,yy}} - {\tilde \gamma_{1,xy}} 
             \over  {\tilde \gamma_{2,xx}} + {\tilde \gamma_{1,xy}}}
,
\end{equation}
where ${\tilde \gamma_{1,ij}}$ and ${\tilde \gamma_{2,ij}}$ 
are the double partial derivatives of ${\tilde \gamma_1}$ and ${\tilde \gamma_2}$.

The frequency shift in the spectrum of a lensed background galaxy
caused by the gravitational field of a tangentially moving cluster of galaxies 
can be read from the time component of the four-vector $\Delta$ 
(Equation~\ref{E:Delta}, again assuming $\delta \ll 1$ and $ \beta \ll 1$) as

\begin{equation} \label{E:deltanu}
  {\Delta \nu \over \nu_0 } = \gamma \beta \delta_x
,      
\end{equation}
where $\beta = v_T/c$, and $\delta_x$ is the $x$ component of the rest 
frame gravitational bend angle (Equation~\ref{E:restframe}). 
In the case of a gravitational potential with spherical symmetry, this reduces to 

\begin{equation} \label{E:delnu_sphe}
 {\Delta \nu / \nu_0 } = \gamma \beta \delta \cos \varphi
, 
\end{equation}
where $\varphi$ is the angle measured from the direction of the tangential 
velocity of the moving lens, which is identical to the expression derived by 
Birkinshaw \& Gull (1983, their Equation 9) and to first order in $\beta$, 
to Equation (2) of Gurvits \& Mitrofanov (1986).

Since there is a time delay between the two images of the strong lensed 
background galaxy, there is a small additional redshift difference, $\Delta z$,  
between the lines of the two images due to the expansion of the Universe.
This can be approximated as

\begin{equation} \label{E:del_z}
 \Delta z = (1+z_{lens})\;H(z_{lens})\;\Delta t 
, 
\end{equation}
where $H(z_{lens})$ is the Hubble parameter evaluated at the redshift of the lens, 
$z_{lens}$, and $\Delta t$ is the time delay difference between the two images.
However, for a typical cluster, this redshift, 
$\Delta z \approx 5 \times 10^{-11} \;(1+z_{lens})\;(H_0/ 50\; {\rm km\; s^{-1}\; Mpc^{-1}}) (\Delta t/ 1\;{\rm yr})$, 
yields a velocity difference only
of order $20\ \rm m\,s^{-1}$ and is negligibly small. 
The frequency shift in Equation~(\ref{E:deltanu}) is not measurable unless one can find a 
reference frame, or measure differences in frequency shifts
in multiple images of a single background object.

\section{Determining Tangential Velocities}
\label{s:Determining}

%
%

\begin{figure}[t]
\centerline{
\plotone{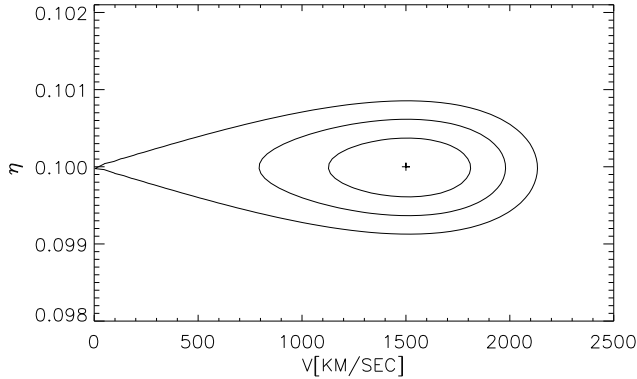}
}
\caption{
$\Delta \chi^2$ contours corresponding to 1, 2, and 3$\sigma$ 
for the elliptical potential of Equation (\ref{E:psi}).
2D cut: ellipticity, $\eta$ and tangential peculiar velocity, $V$. 
The $+$ sign shows the fiducial model parameters, 
$\eta = 0.1$ and $V = 1500\; \rm km\;s^{-1}$
(see text for details of the simulation).
\label{F:FIG1}
}
\end{figure}

%
%

\begin{figure}[t]
\centerline{
\plotone{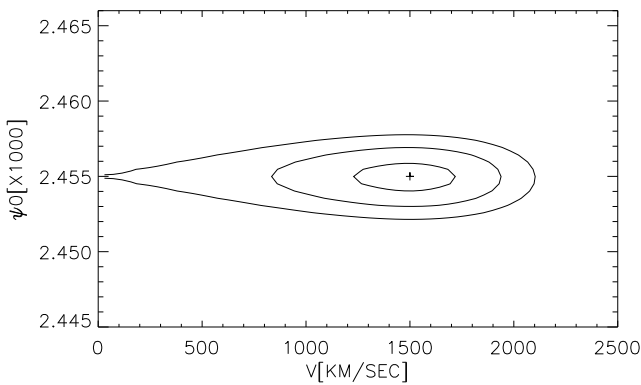}
}
\caption{
$\Delta \chi^2$ contours corresponding to 1, 2, and 3$\sigma$ for an elliptical potential.
2D cut: normalization of the deflecting gravitational potential, $\psi_0$
and tangential peculiar velocity, $V$.
The $+$ sign shows the fiducial model parameters,
$\psi_0 = 2.455 \times 10^{-3}$ and $V = 1500\; \rm km\;s^{-1}$
(see text for details of the simulation).
\label{F:FIG2}
}
\end{figure}

\subsection{Weak Lensing}
\label{ss:Lensing}

We carried out simulations to quantify the observability of the tangential
peculiar motion on the shear field as revealed by images of
background galaxies (Equations~\ref{E:gamma1} and 12). 
The shear parameter, $\tilde{\bfgamma}$, 
is directly related to the measurable ellipticity parameter, ${\bf e}=(e_1, e_2)$, 
$\tilde{\bfgamma} \approx {\bf e}$, since $\kappa \ll 1$. 
The ellipticity parameter as a function of position 
can be directly measured from the images of lensed galaxies
using their isophotes (e.g., Schneider et al. 1992). 
We assumed that ellipticities could be measured for 40 galaxies
per arcmin$^2$, over an area of $50 \times 50$ arcmin$^2$, 
and that the ellipticity parameters of the background galaxies
can be measured with 20$\%$ error. We average over 0.5 arcmin$^2$ cells, containing
20 galaxies each to obtain 5$\%$ ellipticity measurements in $N=5000$ samples.
We choose an elliptical gravitational deflection potential, $\psi$, which approximates 
the potential of a singular isothermal mass distribution at large radii
(see for example Schneider and Bartelmann 1997)

\begin{equation}  \label{E:psi}
  \psi = \psi_0 \Biggl[r_{core}^2+{x^2\over (1-\eta)^2}+{y^2\over (1+\eta)^2}\Biggr]^{1\over 2}
,
\end{equation}
with $\eta = 0.1$, $r_{core} = 1 \arcmin$, and $\psi_0 = 2.455 \times 10^{-3}$ to show
that the shape parameters of the potential ($\psi_0$,$r_{core}$,$\eta$) and the 
tangential peculiar velocity of the cluster decouple, i.e. can be determined separately.
We evaluated the errors using the $\Delta \chi^2$ statistic, where we minimize 

\begin{equation}  \label{E:S}
 S(\beta,\psi_0,r_{core},\eta) = \sum_{i=1}^{i=N} \sum_{j=1}^{j=2} 
          \;{ \bigl[ {\tilde \gamma}_j(x_i,y_i) - e_j(x_i,y_i) \bigr]^2\over \sigma^2_{ij}}
.
\end{equation}
$\tilde{\bfgamma}$ is calculated using Equations~(\ref{E:gamma1}) and 
(13), and the ellipticity parameters, $\bf e$ are obtained by means of Monte Carlo 
simulations assuming errors $\sigma_{ij}$. 
$S$ should be distributed like $\Delta \chi^2$ with $2 N-4$ degrees of freedom.
Our simulations show that $S$ is a non-degenerate four-dimensional ellipsoid near its
minimum, so that $\beta$, $\psi_0$, $r_{core}$, and $\eta$ are independently determined,
and that a tangential peculiar velocity of ${\bf v}_T = 1500$ $\rm km\,s^{-1}$ can be 
measured to $\pm$~300 $\rm km\,s^{-1}$ (1$\sigma$,
see Figures~\ref{F:FIG1} and \ref{F:FIG2}).

\subsection{Frequency Shift}
\label{ss:Frequency}

The method of measuring the tangential peculiar velocity of a cluster of
galaxies proposed by Birkinshaw \& Gull (1983) and corrected by
Gurvits \& Mitrofanov (1986) and Pyne \& Birkinshaw (1993) relied
on observing the brightness change of the CMBR arising from the 
frequency effect, Equation~(\ref{E:delnu_sphe}). 
However, it would also be possible to measure this frequency effect directly 
if there were a background with a sharp line feature. 
Such a background is available in the case of strongly-lensed multiple 
images of a single background galaxy.

We illustrate this method using the well-studied, rich cluster of galaxies, 
CL~0024+1654 (Shapiro \& Iliev 2000; Tyson et al. 1998; Dressler, Gunn \& Schneider 1985), 
at a redshift of 0.395, which produces multiple images of a 
background galaxy located at a redshift of 1.675 \citep{Broa00}.
The surface mass density of the cluster was found to be

\begin{equation}  \label{E:Sigma}
  \Sigma(r) = K \Biggl[ 1 + \zeta \,{r^2 \over r_{core}^2} \Biggr]
                \Biggl[ 1 + {r^2 \over r_{core}^2} \Biggr]^{\zeta - 2}
,
\end{equation}
with $K=7900$ $M_\odot$ pc$^{-2}$, $r_{core}=35$ kpc, and $\zeta=0.57$ 
\citep{ShapIlie00}.
The deflection angle for a spherical potential can be expressed as 

\begin{equation}  \label{E:delta_r}
\delta(r) = { 4 G {\cal M}(r) \over c^2 r }
, 
\end{equation}
where ${\cal M}$ is the mass included within cylindrical radius, $r$.
The relative frequency shift between arcs in the direction of, and opposite 
to the direction of tangential motion of the cluster becomes
twice as large as the single-image frequency shift in Equation~(\ref{E:deltanu}).
Assuming that the tangential velocity $v_T =$  
\\
$1000 \ \rm km\, s^{-1}$, 
and is aligned with two of the lensed images, we obtain a frequency 
shift in emission (or absorption) lines equivalent to a Doppler shift of 
1 $\rm km\,s^{-1}$. 
To measure this effect we must find narrow lines from the same spatial location 
of the imaged background galaxy in each arc, and measure the frequency difference 
between them. Narrow forbidden emission lines of ${\rm [OIII]}$ at 
$\lambda \lambda$4959, 5007, ${\rm [OII]}$ at $\lambda \lambda$3727, 3729, and 
${\rm [NII]}$ at $\lambda$6584, shifted into the infra-red (IR) band should be suitable.

High signal/noise spectra of such lines should be able to fix
their central wavelengths to a few per cent of their widths. 
The precision of the measurement will be dependent on the
stability of the spectrograph (which is better than 
$1 \ \rm km \, s^{-1}$ for stable systems such as Phoenix on
Gemini South; Hinkle et al. 2000) and the intrinsic widths of
the lines being used. Narrow-line emitting regions within
galaxies should have line widths of under $50 \ \rm km \,
s^{-1}$, and so spectra with signal/noise $10^3$ would be
needed to fix the line centroids precisely enough to measure
the cluster tangential velocity to $1000 \ \rm km \, s^{-1}$
or better. Such a high signal/noise cannot currently be
achieved in the near IR at the magnitudes of interest for
CL~0024$+$1654, but would be possible with 50-m class 
telescopes.

\section{Conclusion}
\label{s:Conclusion}

In this letter we proposed two new methods for measuring tangential peculiar velocities of 
clusters of galaxies. 
Our first method is based on weak gravitational lensing. 
We carried out simulations to estimate how well we could determine tangential velocities 
of clusters from distortions of the shapes of background lensed galaxy images.
We assumed that the ellipticities of background galaxies can be measured with 20$\%$ error.
With today's technology, obtaining such a large-scale and accurately measured shear field
is possible only for a limited area. 
In order to measure this effect one would need to reduce the scatter in the intrinsic 
ellipticities of the background galaxies so that this noise source does not swamp the effect 
of tangential motion. 
One possible way of doing this would be to select the spherical components 
of lensed background galaxies, perhaps observing the bulge components in the 
IR for these galaxies using next generation space telescopes.

Our second method is based on measuring frequency shifts between multiple images 
of a single strongly lensed background galaxy. 
This method allows us to measure components of tangential velocities of clusters 
in the direction of pairs of images of strongly-lensed galaxies by measuring 
their relative redshift. We used cluster CL~0024+1654 to demonstrate that 
the needed velocity resolution is achievable with present-day
spectrographs. However, the signal/noise requirements are not
achievable with existing near-IR instruments and telescopes,
with improvements of a factor $10$ in throughput needed to
make the measurement possible in this band.

We might expect clusters of galaxies situated in
superclusters to show the largest tangential velocities, since
these clusters seem to move faster than "field clusters" \citep{Colbet00}. 
While the differential redshift technique should
still be feasible for such clusters, the shear technique may be
compromised by the presence of the large-scale shear field of the
supercluster.

\acknowledgments

This work (SMM) was partly supported by NASA LTSA Grant NAG5-3432.

%
%


\end{document}